\documentstyle[12pt,graphicx]{article}
\title{ Numeric Spectrum of  Relic Gravitational Waves in  Accelerating Universe }
\author{Yang  Zhang,  Wen  Zhao, Yefei Yuan, Tianyang Xia \\
        Astrophysics Center \\
        University of Science and Technology of China \\
        Hefei, Anhui, China }
 \date{}

\begin{document}
\maketitle
\baselineskip=19truept
\def\vek{\vec{k}}

\newcommand{\be}{\begin{equation}}
\newcommand{\ee}{\end{equation}}
\newcommand{\ba}{\begin{eqnarray}}
\newcommand{\ea}{\end{eqnarray}}

\sf
\small

\begin{center}
\Large  Abstract
\end{center}
\begin{quote}
 {\small
The accelerating expansion of the Universe at the present stage is
a process that will change the spectrum of relic gravitational
waves.
 Here we present a numerical calculation for the
power spectrum of relic gravitational waves in the accelerating
Universe.
 The results show that
 although the overall features of the power spectrum is similar to that
 in the non-accelerating   models,  the amplitude is smaller by an order of $10^{-1}$.
 We also find that the spectrum is very sensitive to the index
 $\beta$ of the inflationary expansion with the scale factor
 $a(\tau) \propto |\tau|^{1+\beta}$.
 With increase  of $\beta$,
the resulting spectrum is tilted to be flatter with  more power on
high frequencies, and the sensitivity of the second science run of
the LIGO detectors puts a restriction on the parameter $\beta <
-1.8$. The influence of reheating  following the inflation has
been examined. }
\end{quote}



e-mail: yzh@ustc.edu.cn

\newpage
%\twocolumn
\baselineskip=19truept

In the past  the stochastic background of relic gravitational
waves has been extensively studied \cite{starobinsky}
\cite{rubakov} \cite{harari}. The details of the spectrum of relic
gravitational waves depends not only on the  early stage of
inflationary expansion, but also on the expansion behavior of the
subsequent stages, including the current accelerating expansion.
The calculations of spectrum  so far \cite{allen} \cite{sahni}
\cite{grishchuk} \cite{riazuelo} \cite{tashiro} \cite{henriques}
have been done for the transitions from the inflationary era  to
the radiation,  or  matter dominated state, both  being  a
decelerating expansion. However, the recent observations
\cite{riess} \cite{perlmutter} \cite{spergel} indicate that the
Universe  is currently under accelerating expansion, which may be
driven by the cosmic dark energy ($\Omega_{\Lambda} \sim 0.7$)
plus the dark matter ($\Omega_{m} \sim 0.3$)
   with $\Omega =1$ \cite{zhy}.
The  current accelerating expansion  of Universe will have an
impact on the relic gravitational waves. We have studied this
effect and obtained an approximate  spectrum of the relic
gravitational waves using an analytic method \cite{zh}. In this
paper we present a numerical calculation of the spectrum. The
resulting spectrum from the numerical calculation agrees with that
from the analytic one. Throughout the paper we use units with
$c=\hbar=1$ and adopt notations similar to that of Grishchuk
\cite{grishchuk} for convenience for comparison.

The history of overall expansion of the Universe can be  modelled
as the following sequence of successive epochs of power-law
expansion.

The initial stage (inflationary) \be a(\tau ) = l_0 \mid {\tau}
\mid ^{1+\beta},  \,\,\,\,\, -\infty < \tau \leq  \tau_1, \ee
where  $\tau $ is the conformal  time, $1+\beta<0$, and
$\tau_1<0$. The special case of $\beta=-2$ is the de Sitter
expansion.

The  z-stage \be a(\tau) = a_z(\tau-  \tau_p)^{1+\beta_s},
\,\,\,\,\,  \tau_1 \leq  \tau \leq  \tau_s. \ee The reheating
process towards the end of inflation can be quite complicated and
is model-dependent. This z-stage for various $\beta_s$ can
represent a general reheating epoch .

The radiation-dominated stage \be a(\tau) = a_e(\tau  -\tau_e),
\,\,\,\,\, \tau_s \leq  \tau \leq  \tau_2. \ee

The matter-dominated stage \be a(\tau) =  a_m(\tau  -\tau_m)^2 ,
\,\,\,\,\,  \tau_2  \leq  \tau \leq  \tau_E, \ee where $\tau_E$ is
the time when the dark energy density $\rho_{\Lambda}$ is equal to
the matter energy density $\rho_m$. Before the discovery of the
accelerating expansion of the Universe, the current expansion was
usually taken to be in this matter-dominated stage, which is a
decelerating one. The value of the redshift $z_E$ at the time
$\tau_E$ is given by $ (1+z_E) = a(\tau_H)/a(\tau_E) $, where
$\tau_H$ is the present time. Since $\rho_{\Lambda}$ is constant
and $\rho_m(\tau) \propto a^{-3}(\tau)$,
 one has
$
1=\frac{\rho_{\Lambda}}{\rho_m(\tau_E)}=\frac{\rho_{\Lambda}}{\rho_m(\tau_H)(1+z_E)^3}.
$ If the current values $\Omega_{\Lambda} \sim  0.7$ and $\Omega_m
\sim  0.3$ are taken, then it follows that $ 1+z_E =
(\frac{\Omega_{\Lambda}}{\Omega_m})^{1/3} \sim 1.33. $

The accelerating stage (up to the present) \be a(\tau) =  l_H (
\tau_{a}-  \tau) ^{-1}, \,\,\,\,\,  \tau_E \leq  \tau  \leq \tau_H
. \ee This stage describes the accelerating expansion of the
Universe, which is the new feature in our model and will induce
some modifications to the spectrum of the relic gravitational
waves. It should be mentioned that Eq.(5) is an approximation to
the current expansion behavior since the matter component also
exists in the current Universe.

Given $a(\tau)$ for the various epochs, the derivative
$a'=da/d\tau$ and the ratio $a'/a$ follow immediately. There are
eleven constants in the above expressions of $a(\tau)$, among
which the $\beta$ is imposed upon as the inflationary model
parameter. By the continuity conditions  of $a(\tau)$ and
$a(\tau)'$ at the  four given joining points $\tau_1$, $\tau_s$,
$\tau_2$, and $\tau_E$, one can fix only eight constants. Besides,
choosing  the overall normalization of $a$ at the present time
$\tau_H$ and taking the expansion rate as the observed Hubble
constant will fix  two other constants. Specifically,  we put $(
\tau_a  -   \tau_H ) = 1$ as the normalization of $a(\tau)$ at the
present time $\tau_H$, which fixes the constant $\tau_a$,
 and the constant $l_H  $ is fixed by the following calculation
\be \frac{1}{H} \equiv  \left(\frac{a^2}{a'}   \right)_{\tau_H} =
l_H  , \ee so $l_H$ turns out to be just the  Hubble radius  at
present. Thus after a value of the model parameter $\beta$ is
picked up, all the eleven constants in $a(\tau)$ are fixed up. In
the expanding Robertson-Walker spacetime the physical wave length
is related to the comoving wave number by \be \lambda \equiv
\frac{2\pi a(\tau)}{k}, \ee and the wave number $k_H$
corresponding to the present Hubble radius is \be
 k_H = \frac{2\pi a(\tau_H )}{l_H}  =2\pi .
\ee

Incorporating the perturbations to the spatially flat
Robertson-Walker space-time, the metric is \be ds^2=a^2(\tau) [
d\tau^2-(\delta_{ij}+h_{ij})dx^idx^j ], \ee where  the
perturbations of spacetime $h_{ij}$ is a $3\times 3$ symmetric
matrix.
 The gravitational wave field is
the tensorial portion of $h_{ij}$, which is transverse-traceless
$\partial_i h^{ij}=0$, $\delta^{ij}h_{ij}=0.$ Since  the relic
gravitational waves are very weak,
 $|h_{ij}|  \ll 1$,
so one need just  study the linearized field equation: \be
\partial_{\mu}(\sqrt{-g}\partial^{\mu}h_{ij}({\bf{x}} ,\tau))=0 .
\ee

For  a fixed wave number $\bf k$ and a fixed polarization state
$\lambda$, the wave equation reduces to the second-order ordinary
differential  equation \cite{zh} \cite{grish} \be
 h_k^{(\lambda)''}(\tau)
+2\frac{a'}{a}h_k^{(\lambda)'} (\tau)+k^2h^{(\lambda)} _k =0, \ee
 where the prime denotes $d/d\tau$. Since the equation of
 $h_{\bf k}^{(\lambda)} (\tau) $
 for each polarization is the same, we denote $h_{\bf k}^{(\lambda)}
(\tau) $ by $h_{\bf k}(\tau) $ in below.

The power spectrum $h( k, \tau)$ of  relic  gravitational waves is
defined by the following equation \be \int^{\infty}_0
h^2(k,\tau)\frac{dk}{k}  \equiv <0|h^{ij} ( {\bf x},\tau) h_{ij}(
{\bf x},\tau) |0>, \ee where the right-hand-side is the vacuum
expectation value of the  operator $ h^{ij} h_{ij} $. Taking the
contribution from each polarization to be the same, one reads the
power spectrum \be h(k,\tau) =
\frac{4l_{Pl}}{\sqrt{\pi}}k|h_k(\tau)|. \ee Once the mode function
$h_k(\tau)$ is given, the spectrum $h(k,\tau)$ follows.

The initial condition is taken to be during the inflationary
stage. For a given wave number $k$, its wave  crossed over the
horizon at a time  $\tau_i$, i.e. when the wave length $\lambda_i
= 2\pi a(\tau_i)/k$ is equal to $1/H(\tau_i)$, the Hubble radius
at time $\tau_i$. Eq.(1) yields $1/H(\tau_i) = l_0
|\tau_i|^{2+\beta}/|1+\beta|$, and for the exact de Sitter
expansion with $\beta =-2$ one has $H(\tau_i)=l_0$. Note that a
different $k$ corresponds to a different time $\tau_i =
2\pi|1+\beta|/k$. Now choose the initial amplitude of  mode
function $h_k(\tau)$ as \be |h_k(\tau_i)| = \frac{1}{a(\tau_i)}.
\ee Then the initial  amplitude of the power spectrum is \be h(k,
\tau_i) = A(\frac{k}{k_H})^{2+\beta} ,
  \ee where the constant
   \be
A=  \frac{8\sqrt{\pi}}{|1+\beta|^{1+\beta }} \frac{l_{Pl}}{l_0}.
\ee The power spectrum for the primordial perturbations of  energy
density is $P(k)\propto |h( k, \tau_H)|^2 $, and its spectral
index $n$ is defined as $P(k) \propto  k^{n-1}$. Thus one reads
off the relation $n= 2\beta +5$.
 The exact de Sitter  expansion with $\beta = -2$ will
yield the so-called scale-invariant spectral index $n=1$.

The spectral  energy density parameter $\Omega_g(\nu) $ of the
gravitational waves is defined through the relation $\rho_g/\rho_c
=\int \Omega_g(\nu) \frac{d\nu}{\nu}$, where $\rho_g =
\frac{1}{32\pi G} h_{ij,\, 0} h^{ij}_{\,\, ,\, 0} $ is the energy
density of the gravitational waves, and $\rho_c$ is the critical
energy density. One reads
\[
 \Omega_g(\nu) = \frac{\pi^2}{3}h^2(k, \tau_H)(\frac{\nu}{\nu_H})^2   .
 \]

The spectrum $h(k,\tau_H)$ has an overall factor $A$, which is
fixed as follows. If  the CMB anisotropies at low multipoles are
induced by the gravitational waves, or,  if the contributions from
the gravitational waves and from the density perturbations are of
the same order of magnitude,
 we may assume $\Delta T/T \simeq h(k, \tau_H)$.
The observed CMB anisotropies \cite{spergel} at lower multipoles
is $\Delta T/T \simeq 0.37 \times 10^{-5}$ at $l \sim 2$. Taking
this to be the perturbations at the Hubble radius $1/H$ yields \be
h(k_H, \tau_H) = A \frac{1}{ (1+z_E)^3 }    = 0.37  \times
10^{-5}. \ee

To facilitate our numerical computation, we introduce a time
variable $t$ by \be e^{t}= |\tau-\tau_{0}| ,
 \ee
then the wave equation of $k$-mode is written as\
 \be
\ddot{h}_{k}+(\frac{2\dot{a}}{a}-1)\dot{h}_{k}+k^{2} e^{2t}h_k=0,
\ee where the dot denotes  $d/dt$. This is a second order
differential equation with the function  $\dot{a}/a$ occurring in
the coefficient of $\dot{h}$ being given piecewise for each
expansion stage. For each given $k$ the equation  can be solved by
using the Runge-Kutta method, yielding the $k$-mode wave function
$h_k(t)$. We have done all this for various values of $\log_{10} k
$ in the range $(-25, 20)$, in each time  we take an increase
$\Delta\log k = 0.1$. Switching back to the conformal time $\tau$
as given in Eq.(19) we arrive at $h_k(\tau)$ as well as
$h(k,\tau)$ through Eq.(13). Collecting all the values of
$h(k,\tau)$ at the fixed time  $\tau=\tau_H$ yields the spectrum
function $h(k,\tau_H)$ with $k$ being its variable now.
 In the numeric calculation the following specifications are made.

 The five time instants
  $\tau_{1},\tau_{s},\tau_{2},\tau_{E}$, and
$ \tau_{H}$ determine the overall expansion of the Universe in
various stages, and they must be given as the parameters of our
cosmological
 model.
 In this paper we fix them by the following equations:
 \be
a(\tau_H)/a(\tau_E)= 1.33,
 \ee
 \be
 a(\tau_E)/a(\tau_2)=3454,
 \ee
 \be
 a(\tau_2)/a(\tau_s)=10^{24},
 \ee
 \be
 a(\tau_{s})/a(\tau_1)=300.
 \ee
The explanation to these relations are given as follows.
 Here we assume that $\Omega_{\Lambda} =0.7$ and $\Omega_m =0.3$, and
 that $a(\tau_H)/a(\tau_E)= (1+z_E)= (\Omega_{\Lambda} /\Omega_m )^{1/3} \simeq 1.33$.
 The time instant $\tau_2$ is the moment when
 the matter domination started off,
and the WMAP observation has indicated that the corresponding
red-shift is  $(1+ z_2) \simeq 3454$, so the second relation above
follows. For the radiation dominated stage  from $\tau_s$ to
$\tau_2$, we assume that the starting temperature  is $T_s \simeq
10^{15}$ Gev, a typical energy scale of grand unified theories,
 and that the ending temperature is $T_2 \simeq 1$ ev.
 Then from $a(\tau_2)/a(\tau_s)= T_s/T_2$ the third relation follows.
The $z-$ stage from $\tau_1$ to $\tau_s$ represents the reheating
of the early Universe, which has not been properly understood so
far by cosmologists. For definiteness in numerical computation, we
have chosen the relation $a(\tau_{s})/a(\tau_1)=300$, whose
implications are not in conflict with the observations on relic
gravitational waves. The other nine  constants are $l_0$, $a_z$,
$a_e$, $a_m$, $\tau_{p}$, $ \tau_{e}$, $ \tau_{m}$ ,  $\beta_s$,
and $\beta$, which are  fixed by smoothly joining
 $a(\tau)$ and $a'/a(\tau)$  at the four jointing points,
 giving a relation between  $\beta_s$ and  $\beta$.
For the primordial perturbation spectrum to be close to the
scale-invariant one, we take $\beta=-1.8,~-1.9,~-2.0$, and  obtain
$\beta_{s}=0.598,~-0.552,~-1.689$,  respectively. The initial
condition  is taken
 at the horizon-crossing during the  i-stage of inflation,
 and it consists of the initial values of $h_k(\tau_i)$ and $h'_k(\tau_i)$ as well.
We have already specified $h_k(\tau_i)$ as is implied in Eq.(15),
and the other is taken to be   $h'_k( \tau_i )=0$, which reflects
the fact that $h_k(\tau)$ is a constant outside the horizon $1/H$.

 The resulting  spectrum of relic gravitational waves
 is presented  in Fig.1 for the accelerating Universe,
 and in Fig.2 for the matter-dominated
 non-accelerating Universe.
 These spectra have an oscillating behavior caused by the
 Bessel functions contained in $h_k(\tau)$ as has been expected.
 We also plotted in  Fig.3 their fitting curves,
 which are smoother and reflect the corresponding spectral amplitudes .
 As the figures  show, for a given  $\beta$,
 the amplitude is smaller by a factor $\sim 10^{-1}$
 in the accelerating Universe than in the non-accelerating Universe.
 The spectrum depends on $\beta$, and
a small  $\beta$ gives a small spectrum amplitude of gravitational
waves. The recent second  science run of the LIGO interferometric
detectors \cite{r-abbott} gives a best sensitivity  $3\times
10^{-24}$ near a frequency $\sim 300$Hz. Our  calculation for the
$\beta=-1.9$ case yields  an amplitude $h \simeq10^{-26}$,  much
smaller than the sensitivity. However, in the  model of $\beta
=-1.8$, the amplitude  of the gravitational waves is just to fall
into the sensitivity. Since the second  run of LIGO  has not yet
observed any signal of stochastic gravitational waves in this
frequency range, we arrive at a constraint $ \beta <  -1.8 $
 on the inflationary model.
 Although   relic gravitational waves
 are still difficult to detect directly  at present,
 future observations on CMB polarizations may reveal some information on it \cite{zhh}.
Fig. 4 shows the effects of the existence of reheating $z$-stage
during very early universe, which tends to increase the amplitude
of relic gravitational waves in the high frequencies
$\nu>10^{8}Hz$. In the range  $\nu<10^{8}Hz$, both spectra are
almost the same.

In summary, we have presented a numerical spectrum of  relic
gravitational waves in the present accelerating Universe. The
result confirms our approximate analytic one. In the  frequency
range $ \nu > 10^{-20}$Hz the spectral amplitude is smaller by
$\sim 10^{-1}$ as compared with the non-accelerating model. A
larger value of $\beta$ yields  a flatter spectrum with more power
on the higher frequencies. The resulting sensitivity of  the  LIGO
detectors has put a restriction on the model parameter $\beta <
-1.8$.

ACKNOWLEDGMENT: We like to thank Prof. T.Y. Huang at Nanjing
University and Profs. X.M. Zhang and C.G. Huang at Institute of
High Energy Physics for stimulating discussions. Y. Zhang's work
has been supported by the Chinese NSF (10173008) and by NKBRSF
(G19990754). Y.F. Yuan is supported by the Special Funds for Major
State Research Projects. W.Zhao's work is partially supported by
Graduate Student Research Funding from USTC.

%\newpage

\newpage

\begin{figure}
\centerline{\includegraphics[width=11cm]{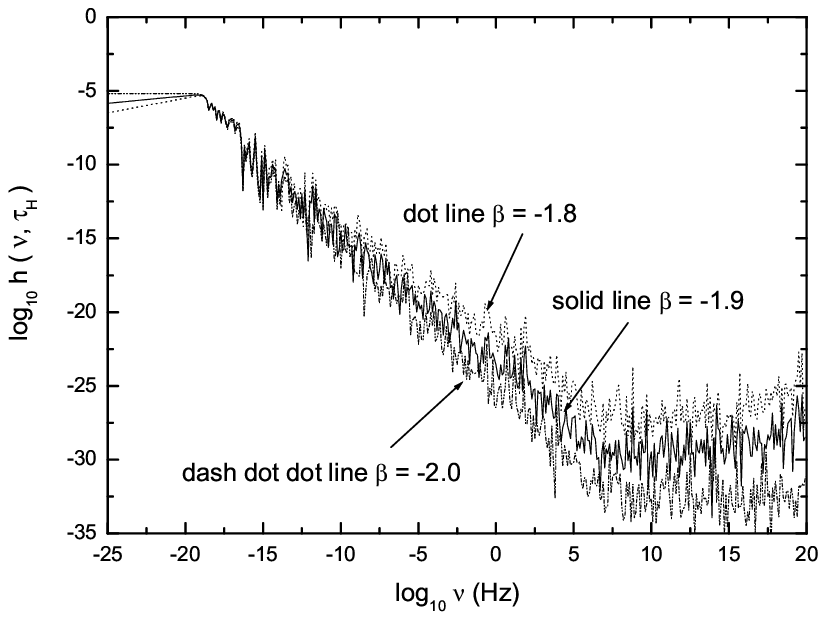}} \caption{
The numerical spectra $h(k,\tau_H)$ in the accelerating universe.
The top (dot), middle (solid), bottom (dash dot dot) lines
correspond $\beta=-1.8, -1.9$ and $-2.0$, respectively. }
\end{figure}

\begin{figure}
\centerline{\includegraphics[width=11cm]{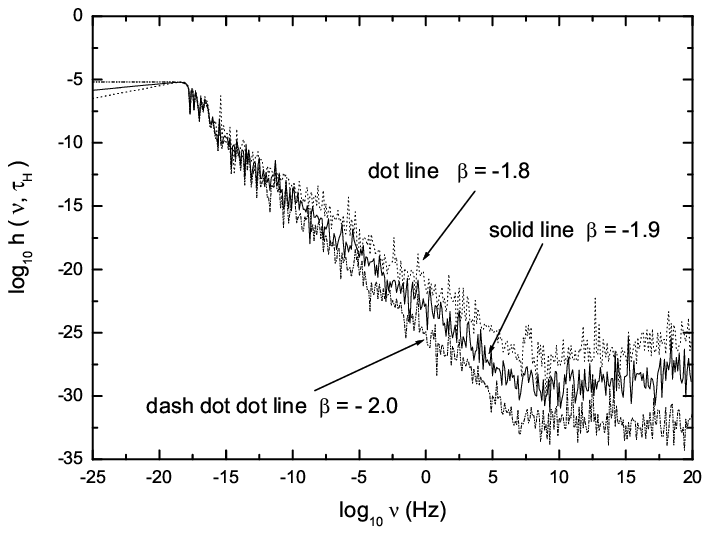}} \caption{
The numerical spectra $h(k,\tau_H)$ in the non-accelerating
universe. The top (dot), middle (solid), bottom (dash dot dot)
lines correspond  $\beta=-1.8, -1.9$ and $-2.0$, respectively.
 }
\end{figure}

\begin{figure}
\centerline{\includegraphics[width=11cm]{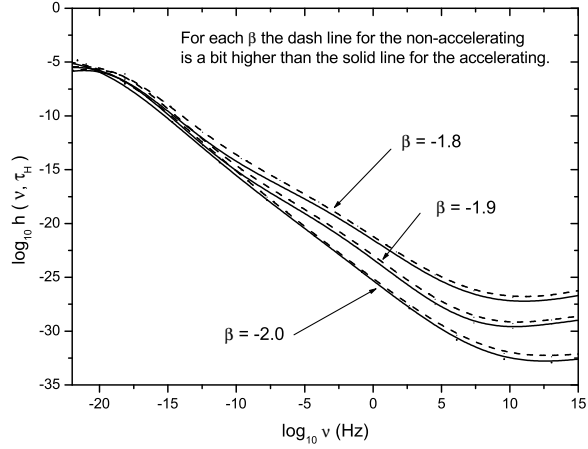}} \caption{
The fitting curves of the numerical spectra $h(k,\tau_H)$. The two
curves on top correspond to  $\beta= -1.8$, the two curves in
middle are $\beta= -1.9$,  and the two curves on bottom are
$\beta= -2.0$. For each fixed $\beta$ the dash  curve is for the
non-accelerating model, and the solid is for the accelerating
model. }
\end{figure}

\begin{figure}
\centerline{\includegraphics[width=11cm]{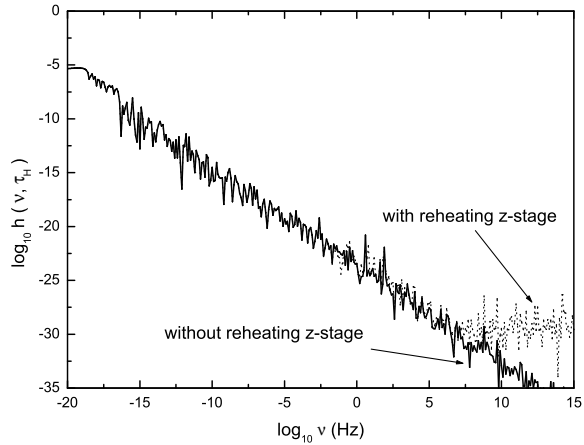}} \caption{
The influence of reheating $z$-stage on the spectra $h(k,\tau_H)$
in the accelerating universe for the $\beta=-1.9$ case. The two
models with or without the $z$-stage give almost the same spectra,
except that at very high frequencies $\nu >  10^{8}$ Hz the
existence of a reheating $z$-stage enhances the amplitude.
 }
\end{figure}

\end{document}